\documentclass[sigconf, screen]{acmart}
\AtBeginDocument{%
  }

\usepackage{array}
\usepackage{subfigure}
\usepackage{graphicx}
\usepackage[switch]{lineno}
\usepackage{amsmath}
\usepackage{amsthm}
\usepackage{multirow}
\usepackage{booktabs}
\usepackage{bbding}
\usepackage{algorithm}
\usepackage{algorithmic}
\usepackage{xcolor}
\usepackage{xspace}
\usepackage{arydshln}
\usepackage[inline]{enumitem}
\usepackage{multicol}

\newcommand{\ie}{{i.e.,}\xspace}

\newcommand{\eg}{{e.g.,}\xspace}
\newcommand{\baby}{{IAM}\xspace}
\newcommand{\babyx}{{IAM}}

\newcommand{\fig}{Figure\xspace}
\usepackage[inline]{enumitem}

\makeatletter

\newcommand{\Rmnum}[1]{\expandafter\@slowromancap\romannumeral #1@}
\makeatother

\setcopyright{cc}
\setcctype{by-nc-nd}
\copyrightyear{2026}
\acmYear{2026}
\acmDOI{XXXXXXX.XXXXXXX}
\acmConference[XXX '26] {Proceedings of xxxx}{April 13--17, 2026}{XXXX, XXXX.}
\acmBooktitle{Proceedings of xxxxxxxxx}
\acmISBN{xxxx/2026/xx}

\begin{document}

\title[From Token to Item: Enhancing Large Language Models for Recommendation \\ via Item-aware Attention Mechanism]{From Token to Item: Enhancing Large Language Models for Recommendation via Item-aware Attention Mechanism}
\author{Xiaokun Zhang}
\affiliation{%
  \institution{City University of Hong Kong}
  \city{Hong Kong}
  \country{China}
}
\email{dawnkun1993@gmail.com}

\author{Bowei He}
\affiliation{%
  \institution{City University of Hong Kong}
  \city{Hong Kong}
  \country{China}
  }
\email{boweihe2-c@my.cityu.edu.hk}

\author{Jiamin Chen}
\affiliation{%
  \institution{City University of Hong Kong}
  \city{Hong Kong}
  \country{China}
  }
\email{jmchen26-c@my.cityu.edu.hk}

\author{Ziqiang Cui}
\affiliation{%
  \institution{City University of Hong Kong}
  \city{Hong Kong}
  \country{China}
  }
\email{ziqiang.cui@my.cityu.edu.hk}

\author{Chen Ma}
\authornote{Corresponding author.}
\affiliation{%
  \institution{City University of Hong Kong}
  \city{Hong Kong}
  \country{China}
  }
\email{chenma@cityu.edu.hk}

\renewcommand{\shortauthors}{Xiaokun Zhang, Bowei He, Jiamin Chen, Ziqiang Cui, and Chen Ma}


\begin{abstract}
Large Language Models (LLMs) have recently gained increasing attention in the field of recommendation. Existing LLM-based methods typically represent items as token sequences, and apply attention layers on these tokens to generate recommendations. However, by inheriting the standard attention mechanism, these methods focus on modeling token-level relations. This token-centric focus overlooks the item as the fundamental unit of recommendation, preventing existing methods from effectively capturing collaborative relations at the item level. 

In this work, we revisit the role of tokens in LLM-driven recommendation and categorize their relations into two types: (1) intra-item token relations, which present the content semantics of an item, \eg name, color, and size; and (2) inter-item token relations, which encode collaborative relations across items. Building on these insights, we propose a novel framework with an item-aware attention mechanism (\baby) to enhance LLMs for recommendation. Specifically, \baby devises two complementary attention layers: (1) an intra-item attention layer, which restricts attention to tokens within the same item, modeling item content semantics; and (2) an inter-item attention layer, which attends exclusively to token relations across items, capturing item collaborative relations. Through this stacked design, \baby explicitly emphasizes items as the fundamental units in recommendation, enabling LLMs to effectively exploit item-level collaborative relations. Extensive experiments on several public datasets demonstrate the effectiveness of \baby in enhancing LLMs for personalized recommendation.
\end{abstract}

\ccsdesc[500]{Information systems~Recommender systems}

\keywords{Recommender Systems, Large Language Models, Item-aware Attention, Collaborative information.}

\maketitle

\section{Introduction}
Sequential recommendation (SR) aims to predict a user’s next interaction based on her historical behavior~\cite{GRU4Rec, SASRec}. By modeling dynamics of user preferences, SR plays a crucial role in real-world scenarios such as e-commerce~\cite{BERT4Rec}, social platforms~\cite{FineRec}, and multimedia streaming~\cite{IP2}. A fundamental driver of existing SR methods is \textbf{collaborative information}, which reflects certain patterns embedded in user-item interactions, \eg co-occurrence relations between items~\cite{Zhang@WSDM2026, Kim@KDD2024}. Mining these collaborative signals from massive user-item interactions enables models to uncover shared interests among users and deliver relevant recommendations.

\begin{figure}[t]
  \centering
  \includegraphics[width=0.90\linewidth]{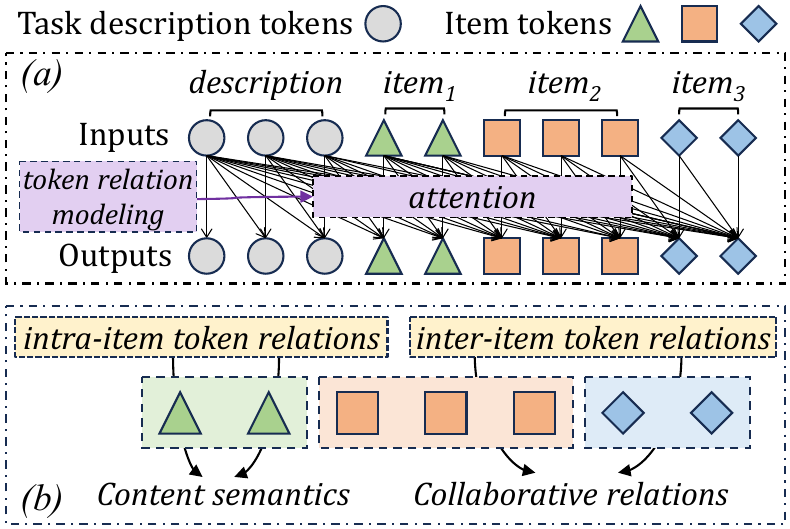}
  \caption{(a) Current LLM-based methods focus on modeling token-level relations, while failing to exploit collaborative relations at the item level; (b) intra- and inter-item token relations indicate item content semantics and collaborative relations, respectively. }\label{intro}
  \vspace{-0.15in}
\end{figure}

Large language Models (LLMs), with their rich world knowledge and strong reasoning capabilities, have recently gained increasing attention in the recommendation research~\cite{Liu@AAAI2025, Ren@WWW2024, Li@RecSys2025}. 
Typically, existing LLM-based methods concatenate task description text with the sequence of item content (\eg titles) as the instruction, convert these texts into a token sequence, and apply attention mechanisms (\eg Transformer architectures) on these tokens to generate recommendations. To integrate collaborative information, these methods generally encode user–item interaction patterns into collaborative tokens for each item. Such tokens could be derived from heuristic item indexing rules~\cite{Hua@SIGIRAP2025, Hou@ICML2025}, or represented as dense vectors learned by well-established neural models (\eg SASRec~\cite{SASRec})~\cite{Kim@KDD2024, ZhangY@TKDE2025}. These collaborative tokens are then concatenated with content tokens from items, guiding LLMs to generate recommendations.

However, these methods suffer from a critical limitation in their design. As shown in~\fig~\ref{intro} (a), by framing recommendation as the processing of token sequences, they explicitly treat tokens as the fundamental modeling unit. \textit{This token-centric paradigm restricts LLM-based methods to attending only to token-token relations, preventing them from recognizing items as independent units within the recommendation context.} Since collaborative relations inherently emerge between items rather than tokens, such methods fail to effectively exploit collaborative information, leading to a critical misalignment with the core objective of recommendation.

To this end, this work revisits the role of tokens in LLM-based recommendation from an item-centric perspective. As shown in~\fig~\ref{intro} (b), token relations can be divided into two categories: intra-item and inter-item. \textbf{Intra-item token relations} capture dependencies among tokens within a single item. These intra-item tokens jointly represent an item's content semantics, such as its name, color, and size. In contrast, \textbf{inter-item token relations} occur between tokens belonging to different items. The interactions of these tokens reflect item-item relations and provide a pathway to uncover collaborative information under recommendation scenarios.

Motivated by these insights, we introduce an \underline{I}tem-aware \underline{A}ttention \underline{M}echanism (\baby) to align LLMs with the recommendation task. Unlike conventional attention that treats all tokens uniformly, \baby explicitly distinguishes between intra- and inter-item token relations through two dedicated layers: an intra-item attention layer and an inter-item attention layer. \textbf{The intra-item attention layer} aims to handle intra-item token relations. Specifically, it models token relations within each item while ignoring cross-item token interactions, enabling precise modeling of item content semantics and reinforcing items as independent units in the task. Complementarily, \textbf{the inter-item attention layer} attends to inter-item token relations. It attends exclusively to token relations across items while discarding within-item interactions, explicitly capturing item-item relations. By stacking these two layers, \baby empowers LLMs to effectively exploit item-level collaborative information and deliver accurate recommendations. In summary, main contributions of this work are as follows,

\begin{itemize}
    \item \textbf{Problem diagnosis}. We identify a critical limitation in current LLM-based recommendation: their token-centric paradigm overlooks the role of items as independent units, leaving them structurally incapable of modeling collaborative information at the item level. To our best knowledge, this work marks the pioneering attempt to shift the focus of LLM-based recommendation from token to item.
    \item \textbf{Methodology}. We propose an item-aware attention mechanism (\baby) tailored for recommendation within LLMs. By distinguishing between intra- and inter-item token relations, \baby explicitly preserves items as independent units, enabling LLMs to effectively exploit item-level collaborative information. 
    \item \textbf{Empirical validation}. Extensive experiments conducted on several public datasets demonstrate the superiority of \baby over existing LLM-based approaches, achieving an average improvement of 34.54\% on standard evaluation metrics. 
\end{itemize}

\section{Related Work}

\subsection{Sequential Recommendation}
Sequential recommendation (SR) aims to capture users' evolving preferences based on their historical behaviors, thereby delivering personalized suggestions~\cite{chen2024shopping, he2024interpretable, MMSBR, Zhang@TKDE2025}. Over the past decade, neural architectures have dominated this task due to their strong representation capacity. A wide spectrum of neural models has been explored to capture collaborative signals from user–item interaction sequences, ranging from Recurrent Neural Networks and their variants~\cite{GRU4Rec, NARM}, attention-based approaches~\cite{STAMP, SASRec}, and Transformer-based methods~\cite{BERT4Rec, Zhang@KDD2023}, to Graph Neural Networks~\cite{SR-GNN, Xu@WWW2025}, and contrastive learning frameworks~\cite{Zhang@PR2026, Cui@CoRR2025}. Beyond sequence-level modeling, several studies partition user interaction sequences into multi-granularity sub-sequences to capture item collaborative relations at different levels~\cite{guo@WSDM2022, Zhang@WSDM2023}. Recently, researchers have explored the integration of multi-modal signals, such as textual and visual content, in an attempt to improve the understanding of item characteristics and user intents~\cite{CoHHN, BiPNet, DIMO, LiYT@RecSys2025M}. Unfortunately, these methods suffer from limited semantic understanding, which constrains their performance.

\subsection{LLM-based Recommendation}
Trained on massive text corpora, Large Language Models (LLMs) have demonstrated remarkable success across a wide range of language tasks~\cite{SETRec, Sun@CIKM2024}. This success has motivated growing efforts to adapt LLMs for recommendation tasks. Early attempts reformulate user–item interactions into natural language, like item title sequences, and combine them with task-specific instructions to fine-tune LLMs for generating recommendations~\cite{P5, TALLRec, Zheng@ICDE2024}. More recent works aim to inject collaborative information directly into LLMs. A common strategy is to create collaborative tokens that encode item collaborative relations for each item and integrate them with item content tokens during fine-tuning. For instance, E4SRec~\cite{E4SRec} simply uses item IDs as collaborative tokens, while other approaches re-index items through heuristic rules for encoding collaborative signals~\cite{Zhu@WWW2024, Hua@SIGIRAP2025, Hou@ICML2025, Liu@SIGIR2025}. Alternatively, methods such as CoLLM~\cite{ZhangY@TKDE2025}, LLaRA~\cite{LLaRA}, A-LLMRec~\cite{Kim@KDD2024}, and iLoRA~\cite{ilora}, leverage well-established collaborative models, \ie SASRec~\cite{SASRec}, to extract item embeddings, which are then treated as collaborative tokens. Despite achieving impressive performance, existing methods still follow the standard attention mechanism of LLMs, which indiscriminately attends to all tokens within the instruction and fails to treat items as independent units. This structural limitation hinders them from effectively capturing relations at the item level, thereby constraining their ability to model collaborative information.

\section{Preliminaries}

\subsection{Problem Statement}
Let $\mathcal{U}$ and $\mathcal{X}$ denote the unique user and item sets, with $m = |\mathcal{U}|$ and $n = |\mathcal{X}|$ denoting the total number of users and items, respectively. For a user $u_i \in \mathcal{U}$, her historical behaviors can be represented as an item sequence $\mathcal{S}$ = [$x_1, x_2, \cdots, x_t$], where each $x_i$ $\in$ $\mathcal{X}$ and $t$ is the sequence length. Each item $x_i$ is associated with some textual features, such as its title, denoted by $r_i$, which consists of a sequence of words. The goal of sequential recommendation is to predict the next item $x_{t+1}$ (\eg the ground truth item) that a user $u_i$ is likely to engage with, based on her previously interacted item sequence $\mathcal{S}$. Important notations used in this work are summarized in Table~\ref{notation} for clarity and ease of reference.

\begin{table}[t]
\tabcolsep 0.08in 
\centering
\caption{Notations and descriptions.}
\begin{tabular}{cl}
\toprule
Notation      & Description\\
\midrule
$x_i$  & an item. \\
$u_i$   & a user. \\
$\mathcal{X}$ & the item set. \\
$\mathcal{U}$  & the user set. \\
$n$ & the total number of items in the item set. \\
$m$  & the total number of users in the user set. \\
$v_i$&  a token derived from item textual features. \\
$\mathbf{e}_i$ &  the embedding of a token $v_i$. \\
$\mathcal{S}$      & an item sequence generated by a user. \\
$\mathcal{S}^{t}$      & the token sequence of $\mathcal{S}$. \\
\multirow{2}*{$ins$}   & an instruction containing both the task description  \\
    & and the item title sequence.\\
$\mathbf{h}_i$ &   the $i$-th output embedding of self-attention. \\
$y_i$ & the predicted score of item $x_i$ to be interacted next. \\
\bottomrule
\end{tabular}

\label{notation}
\end{table}

\subsection{Recommendation Paradigm of LLMs}\label{llmrec}

\begin{figure}[t]
  \centering
  \includegraphics[width=0.92\linewidth]{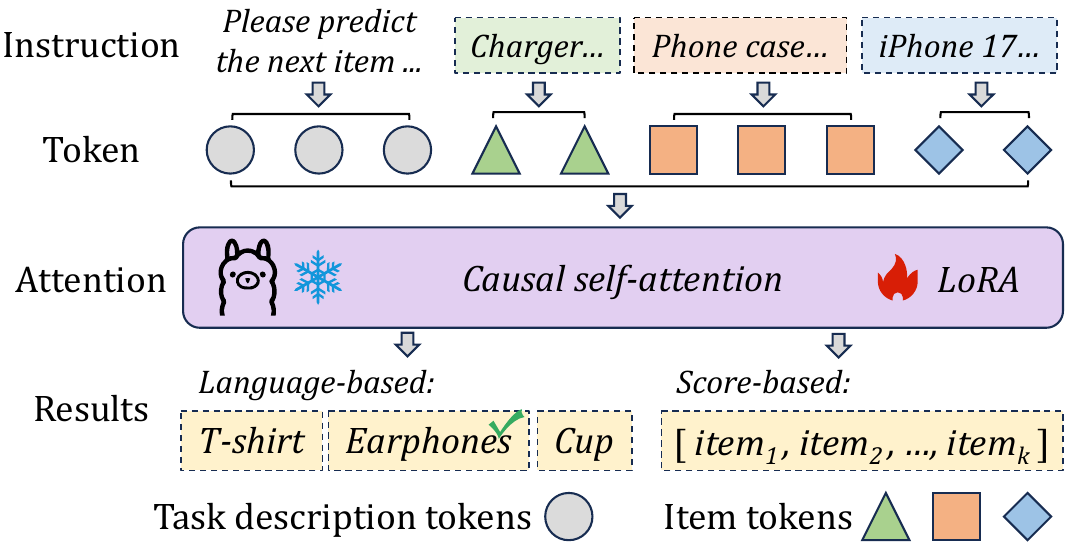}
  \caption{Recommendation paradigm of LLM-based methods.}\label{paradigm}
\end{figure}

Open-source LLMs were originally designed as general-purpose language generators, intended to respond to user queries and support a wide range of tasks. In the context of recommendation, as illustrated in~\fig~\ref{paradigm}, existing studies typically follow a common paradigm to adapt LLMs for this task.

\subsubsection{Instruction construction}
Instruction tuning has emerged as a widely used strategy to unlock the potential of LLMs for specialized tasks~\cite{Ouyang@NIPS2022}. This process involves constructing instruction–response pairs and utilizing them to fine-tune LLMs to perform the downstream task. In the context of recommendation, an instruction $ins$ typically consists of two components: (1) a task description, which defines the recommendation task and guides the LLM to generate the next likely interacted item based on a user's historical interactions; and (2) a sequence of item textual features, \eg item titles [$r_1, r_2, \cdots, r_{t}$], which reformulates the user–item interaction data into natural language. The response, in turn, is the item textual feature, \eg title $r_{t+1}$, of the ground truth item $x_{t+1}$ with which the user interacts next.

\subsubsection{Token layer}
Tokens serve as the basic processing units in LLMs, which may correspond to words, sub-words, or character-level segments~\cite{Juan@ICLR2025}. LLMs are inherently designed to model dependencies among these tokens to achieve language understanding.  In the recommendation task, the textual instruction $ins$ is first decomposed into a token sequence, $\mathcal{S}^{t}$ = [$v_1, v_2, \cdots, v_{l}$], where $v_i$ is a token and $l$ denotes the number of tokens. Each token $v_i$ in $\mathcal{S}^{t}$ is then mapped into a token embedding $\mathbf{e}_i \in \mathbb{R}^{d}$, forming the inputs for LLMs .
To incorporate collaborative information, existing methods often introduce additional collaborative tokens derived from user–item interactions. These collaborative tokens can be constructed in different ways, such as re-indexing items based on their co-occurrence patterns~\cite{Hua@SIGIRAP2025, Liu@SIGIR2025} or precomputing item embeddings using well-established collaborative models~\cite{LLaRA, ilora}. In practice, such collaborative tokens are concatenated with item content tokens, providing LLMs with an augmented input representation that mixes content semantics with collaborative signals.

\subsubsection{Attention layer}
\begin{figure}[t]
  \centering
  \includegraphics[width=0.92\linewidth]{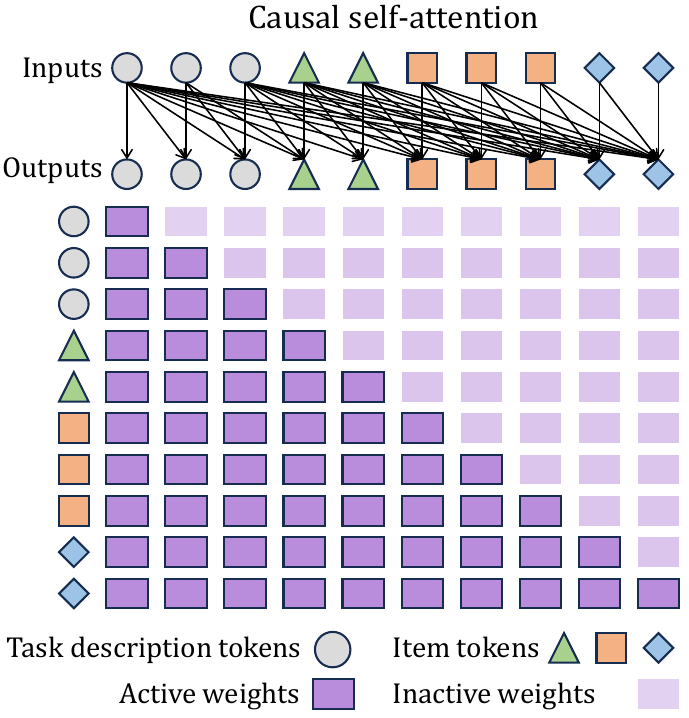}
  \caption{Causal self-attention mechanism and its attention matrix.}\label{causal}
\end{figure}

The core component of LLMs is the causal self-attention mechanism, which models dependencies among tokens while respecting autoregressive constraints~\cite{llama}. Formally, given a sequence of token embeddings $\mathbf{E}$ = [$\mathbf{e}_1, \mathbf{e}_2, \cdots, \mathbf{e}_{l}$] with $\mathbf{e}_i \in \mathbb{R}^{d}$, the output embedding for the $i$-th position, $\mathbf{h}_i \in \mathbb{R}^{d}$, is computed as:
\begin{align}
    \mathbf{h}_i &= \sum_{j=1}^{l}f(\alpha_{ij})\mathbf{V}\mathbf{e}_j,\\
    \alpha_{ij} &= (\mathbf{Q}\mathbf{e}_i)^{\top}(\mathbf{K}\mathbf{e}_j),\\
    f(\alpha_{ij})&=
\begin{cases}
    0, i \textless j,\\
    \alpha_{ij}, i \geq j,
\end{cases}
\end{align}
where $\mathbf{Q}, \mathbf{K}, \mathbf{V} \in \mathbb{R}^{d \times d}$ are learnable projection matrices that map each embedding into the query, key, and value spaces, respectively. The function $f(\cdot)$ denotes the causal mask, which enforces autoregressive constraints by preventing each position from attending to future tokens for language generation. The \fig~\ref{causal} illustrates the working mechanism of causal self-attention, where only the lower-triangular weights of the attention matrix are activated. However, this standard attention mechanism treats all tokens uniformly and concentrates solely on token-level dependencies. Therefore, it fails to perceive the independent item units in the recommendation scenario.

\subsubsection{Recommendation generation}\label{sampling}
In the current literature on LLM-based recommendation, two primary paradigms for generating recommendations can be identified: language-based and score-based methods~\cite{LLaRA, Cui@arXiv2025}. The language-based approach~\cite{LLaRA, ilora} leverages LLMs to directly generate textual features of the next item (typically the item title), aligning with language generation practices in natural language processing (NLP). Due to the vast size of item sets, this paradigm is often simplified by prompting LLMs to select the ground-truth item from a small pre-defined candidate set (commonly around 20 items, including the correct one), which is similar with the negative sampling strategy widely used in sequential recommendation~\cite{he2023dynamically, qiu2024ease}. In contrast, the score-based approach~\cite{E4SRec, Zhu@WWW2024, Xu@ICLR2025} projects the output embeddings of LLMs into a score vector, commonly using a learnable matrix, where each entry indicates the likelihood of a candidate item being the next interaction. The top-k recommendation list can then be derived from these predicted scores. This process is closely aligned with the full-ranking techniques prevalent in neural recommendation models~\cite{NARM, SASRec, BERT4Rec}. 

\subsubsection{Fine-tuning with Low-rank Adaption}
Fully fine-tuning LLMs often demands prohibitive computational resources and storage overhead. To address this challenge, Low-Rank Adaptation (LoRA) has been proposed as a lightweight yet effective alternative technique~\cite{LoRA}. Instead of updating all parameters of the LLM, LoRA injects trainable low-rank matrices into the weight update process of specific layers, \eg attention projections. Following this practice, in the context of recommendation task, researchers rely on LoRA to fine-tune LLMs with instruction–response pairs, where the instruction encodes user-item interaction history and the response corresponds to the next item.

\section{Experimental Setup}

\subsection{Research Questions}

We conduct comprehensive experiments to examine \babyx's performance, with a focus on the following research questions: 

\begin{itemize}
    \item \textbf{RQ1}: How does our \baby perform compared with existing state-of-the-art methods? (ref. Section~\ref{sec:overall})
    
    \item \textbf{RQ2}: How do the proposed intra- and inter-item attention layers, individually and jointly, contribute to the recommendation performance? (ref. Section~\ref{sec:ablation})

    \item \textbf{RQ3}: What is the influence of the key hyper-parameter on \baby? (ref. Section~\ref{sec:hyper})

\end{itemize}

\subsection{Datasets and Preprocessing}

\begin{table}[t]
\tabcolsep 0.12in 
\centering
\caption{Statistics of datasets.}
\begin{tabular}{cccc}
\toprule
Datasets      &Grocery  &Arts  & Cellphones \\
\midrule
\#item        & 1,874   & 4,265   & 6,593    \\
\#user     & 6,025  & 17,432   & 17,639    \\
\#interaction & 44,921  & 134,105 & 114,605  \\
avg.length    & 7.46     & 7.69     & 6.50   \\
\bottomrule
\end{tabular}

\label{statistics}
\end{table}

In this work, three datasets are incorporated to examine the recommendation performance of the proposed \baby and all baseline methods, \ie \textbf{Grocery}, \textbf{Arts} and \textbf{Cellphones}, which are scratched from Amazon\footnote{\url{http://jmcauley.ucsd.edu/data/amazon/}}. These datasets span diverse domains and user interaction patterns, making them representative and popular benchmarks for sequential recommendation~\cite{SASRec, P5, E4SRec, Kim@KDD2024, Hou@ICML2025}. For each dataset, we adopt the item title as textual information for an item to provide its content semantics for LLMs. In line with the common practice~\cite{NARM, LLaRA, ZhangY@TKDE2025}, 5-core filtering is applied for these datasets, where we filter out the items and users with fewer than 5 interactions. Given a user’s interaction sequence, the final item is designated as the prediction label, \ie ground truth item, while the preceding items serve as historical context for modeling user preferences. In addition, we split each dataset chronologically into training, validation, and test sets with an 8:1:1 ratio. The detailed statistics of all datasets are summarized in Table~\ref{statistics}.

\subsection{Evaluation Protocol}
This study adopts the \textit{full-ranking strategy} for sequential recommendation. Specifically, the proposed \baby and all baselines are required to rank the entire item set $\mathcal{X}$ according to the predicted probability of each item being the next interaction. The top-$k$ items with the highest scores form the recommendation list, denoted as $rec$ = [$x_1, x_2, ..., x_k$], where $x_i$ $\in$ $\mathcal{X}$. For evaluation, we follow standard practice~\cite{NARM, SR-GNN, BERT4Rec, P5, LLaRA} and adopt two widely used metrics: \textbf{Prec@k} (Precision), which measures the proportion of test cases in which the ground-truth item appears within the top-$k$ recommendation list; and \textbf{NDCG@k} (Normalized Discounted Cumulative Gain), which considers the rank of the ground truth item among the top-$k$ list. Both metrics are reported at $k = 5$ and $k = 10$, where higher values indicate better recommendation performance.

\subsection{Baseline Methods}
To examine the effect of \baby, we compare it against nine competitive baselines that fall into two categories: traditional neural network-based methods and emerging LLM-based approaches.

Traditional methods: (1) \textbf{GRU4Rec}~\cite{GRU4Rec} leverages Gated Recurrent Units (GRU) to mine sequential patterns between items; (2) \textbf{NARM}~\cite{NARM} augments GRU with an attention mechanism to capture a user’s main intent; (3) \textbf{SASRec}~\cite{SASRec} applies the self-attention mechanism to learn transition dependencies in user behaviors; (4) \textbf{SR-GNN}~\cite{SR-GNN} represents user-item interactions as graphs and employs Graph Neural Networks (GNN) to capture pairwise item transitions; and (5) \textbf{Atten-Mixer}~\cite{Zhang@WSDM2023} partitions a sequence into multiple sub-sequences of varying lengths and applies diverse attention layers to capture fine-grained co-occurrence patterns.

LLM-based methods: (6) \textbf{Llama}~\cite{llama} serves as a representative LLM backbone, fine-tuned with recommendation instructions to enable personalized services; (7) \textbf{P5}~\cite{P5, Hua@SIGIRAP2025} employs various heuristic rules to re-index items for injecting collaborative information into T5 architecture~\cite{T5}; (8) \textbf{E4SRec}~\cite{E4SRec} directly incorporates item ID sequences into instructions, aiming to explicitly integrate collaborative information; and (9) \textbf{LLaRA}~\cite{LLaRA} leverages pre-trained models to generate collaborative embeddings for each item, which are then merged into LLM instructions to explicitly guide the model in preserving collaborative relations among items.

\begin{table*}[ht]
\tabcolsep 0.02in 
  \centering
    \caption{Experimental results (\%) of all methods on three datasets. Bold scores are the best performance, while underlined scores are the second best. Improvements of \baby over the best baseline (*) are statistically significant with $t$-test ($p < 0.01$).}
    \begin{tabular}{c cccc cccc cccc}  
    \toprule  
    \multirow{2}*{Method}& 
    \multicolumn{4}{c}{Grocery}&\multicolumn{4}{c}{Arts}&\multicolumn{4}{c}{Cellphones}\cr  
    \cmidrule(lr){2-5} \cmidrule(lr){6-9} \cmidrule(lr){10-13}
    &Prec@5&NDCG@5&Prec@10&NDCG@10 &Prec@5&NDCG@5&Prec@10&NDCG@10 &Prec@5&NDCG@5&Prec@10&NDCG@10\cr  
    \midrule  
    GRU4Rec     & 10.10& 8.66& 11.66& 9.07      & 28.41& 22.56& 30.40& 23.21    & 3.28& 2.48& 3.78& 2.81 \\
    NARM        & 12.78& 11.38& 13.37& 11.69    & 32.20& 30.14& 33.01& 30.76    & 3.13& 2.53& 4.18& 2.87 \\
    SASRec      & 11.29& 10.39& 13.05& 11.22    & 33.11& 32.07& 34.87& 32.57    & 3.35& 2.60& 4.65& 2.93 \\
    SR-GNN      & 10.63& 9.31& 12.57& 10.22     & 29.37& 28.49& 30.91& 29.65    & 3.60& 1.79& 4.72& 2.19 \\
    Atten-Mixer & 12.82& 11.24& 13.66& 11.81    & \underline{36.05}& \underline{33.92}& \underline{37.81}& \underline{34.62}    & 4.09& \underline{3.06}& 5.15& \underline{3.50} \\
    Llama       & 12.64& 11.37& 13.70& 11.84    & 33.21& 31.33& 34.28& 32.00    & 2.36& 1.89& 4.10& 2.08 \\
    P5          & 12.52& 11.18& 13.48& 11.92    & 35.55& 33.29& 36.68& 33.86    & 3.48& 2.79& 4.84& 3.17  \\    
    E4SRec      & 11.57& 10.71& 13.44& 11.63    & 32.64& 30.69& 33.78& 30.93    & 3.63& 2.74& 5.00& 3.02 \\
    LLaRA       & \underline{13.12}& \underline{11.46}& \underline{13.83}& \underline{12.08}    & 35.24& 33.53& 35.83& 33.96    & \underline{4.31}& 2.87& \underline{5.38}& 3.24 \\
    \midrule
    IAM         & $\bf 14.84^*$ & $\bf 12.38^*$ & $\bf 17.40^*$& $\bf 13.29^*$    & $\bf 37.15^*$& $\bf 34.87^*$& $\bf 40.38^*$& $\bf 35.91^*$    & $\bf 6.37^*$& $\bf 5.13^*$& $\bf 9.20^*$& $\bf 6.12^*$ \\
    $impro.$    & 13.11\%& 8.03\%& 25.81\%& 10.04\%     & 3.05\%& 2.82\%& 6.80\%& 3.74\%        & 47.80\%& 67.65\%& 71.00\%& 74.86\% \\
    \bottomrule
    \end{tabular}
    \label{performance}
\end{table*}

\subsection{Implementation Details}
To ensure a fair comparison, we follow the experimental settings reported in the original papers as closely as possible, aiming to maximize the performance of all baseline methods. The key hyper-parameters for both \baby and baselines are tuned via grid search based on validation performance measured by Prec@10. For neural network-based methods, the embedding dimension is selected from $\{64, 128, 256, 512, 1024, 2048\}$. We adopt a mini-batch size of 512 and optimize models using Adam with an initial learning rate of 0.001. For LLM-based methods, including \baby and baselines, we employ Llama3 with 3B parameters as the default backbone unless otherwise specified. The instruction used to prompt these methods for recommendations is: ``\textit{Please predict the next item a user would purchase, given the following purchased items: <title$_1$>, <title$_2$>, ..., <title$_t$>}''. Fine-tuning is performed using LoRA with a rank of 8, alpha of 16, and a dropout rate of 0.05. In LLaRA, following established practice~\cite{LLaRA, ZhangY@TKDE2025, Kim@KDD2024}, we utilize SASRec~\cite{SASRec} to construct collaborative embeddings for items. For the proposed \baby and all baseline methods, we conduct five independent runs and report the average results as the final performance.

For our \baby, the primary hyper-parameter is the repetition times $q$ of the intra- and inter-item attention layers, which is determined by the depth of the LLM backbone, \eg 16 layers for Llama3-1B, 28 layers for Llama3-3B, and 32 layers for Llama3-8B. 
Notably, \baby represents user behaviors solely through item titles, without the extra efforts to obtain collaborative tokens via heuristic rules or pre-trained models. This design ensures that the performance improvements stem entirely from the proposed attention-layer architecture, which constitutes the core contribution of this work.

\section{Results and Analysis}

\subsection{Overall Performance (RQ1)}\label{sec:overall}

The performance of all baseline methods and the proposed \baby across three datasets is presented in Table~\ref{performance}, from which we can derive the following key observations:

Firstly, among neural network-based methods, Atten-Mixer achieves the best performance in most cases. By partitioning a sequence into multiple sub-sequences and modeling them individually, it captures item-level collaborative relations in a fine-grained manner, which contributes to its superior results. This finding highlights the importance of modeling item-level collaborative signals in the task. In addition, SASRec generally demonstrates competitive performance. Owing to both its effectiveness and parallelizable self-attention mechanism, SASRec has become a widely adopted backbone for encoding collaborative information in recent studies, particularly in LLM-based recommendation frameworks.

Secondly, among the LLM-based methods, the key distinction lies in their manners to incorporate collaborative information, such as re-indexing items via pre-defined rules in P5, direct item ID incorporation in E4SRec, and collaborative embedding injection in LLaRA. The performance differences among these approaches underscore the pivotal role of collaborative information in LLM-based recommendation. This observation also supports the central motivation of this study. In addition, LLaRA consistently outperforms the other LLM-based baselines across all datasets, highlighting the effectiveness of explicitly embedding collaborative information into LLMs for recommendations.

Thirdly, compared with neural models, LLM-based methods generally deliver competitive performance. In particular, Llama, which relies solely on item titles for fine-tuning the LLM, surpasses several representative neural baselines, such as GRU4Rec and NARM, demonstrating the strong potential of LLMs in the recommendation task. Nevertheless, the most recent and advanced Atten-Mixer still outperforms LLM-based methods in certain scenarios. For example, it achieves the best performance among all baselines on the Arts dataset. This suggests that LLM-based approaches still have considerable room for improvement. Furthermore, much of the existing literature tends to benchmark LLMs only against early neural baselines, overlooking comparisons with more advanced models, which may lead to an overestimation of their effectiveness. Our findings thus provide a reminder for both researchers and practitioners to adopt fairer benchmarking practices and critically assess the true capabilities of LLMs in recommendations.

Finally, the proposed \baby consistently outperforms all baseline methods across all datasets and evaluation metrics, demonstrating its strong effectiveness for sequential recommendation. In particular, \baby achieves substantial gains over the best-performing baselines, with improvements in terms of Prec@10 and NDCG@10 by 25.81\% and 10.04\% on Grocery, 6.80\% and 3.74\% on Arts, as well as an impressive 71.00\% and 74.86\% on Cellphones. We attribute this superiority of \baby to its item-aware attention mechanism, which shifts the focus of LLMs from token to item in the task. By distinguishing between intra- and inter-item token relations, this novel attention mechanism drives LLMs to explicitly preserve items as independent units and effectively capture collaborative information at the item level, contributing to providing accurate recommendations accordingly.

\section{Conclusion and Future Work}

In this study, we scrutinize Large Language Model (LLM)-based recommendation approaches and uncover a subtle yet critical limitation in their design: they primarily focus on handling token-token relations, while overlooking the effective capture of collaborative information at the item level. To address this issue, we propose \baby, a novel LLM-based recommendation framework equipped with an item-aware attention mechanism to enhance collaborative information modeling for LLMs. By distinguishing intra- and inter-item token relations through dedicated attention layers, \baby consolidates items as independent units and explicitly exploits collaborative information at the item level. Comprehensive experiments on multiple real-world datasets demonstrate the consistent superiority of \baby over current state-of-the-art methods, including representative neural network-based and LLM-based methods. 

As to future work, we first plan to incorporate extra modality information, like item images, into foundation models to enhance their understanding for item characteristics and user interest. Moreover, current attention layers within LLM-based recommendation methods are uniform across users. It will be a promising direction to design adaptive modules that tailor the attention scope to individual users. In addition, as LLMs can be computationally expensive, designing lightweight variants of \baby with knowledge distillation, quantization, or hybrid architectures would further improve its scalability and implementation for industrial applications.

\newpage

\onecolumn
\begin{multicols}{2}
   \bibliographystyle{ACM-Reference-Format}
   \bibliography{blank_ref}
\end{multicols}


\end{document}